\documentclass{evn2002}
\usepackage{graphicx}
\begin{document}
   \title{Oblique Polarization Structures in AGN and QSO Radio Jets}                    
          
   \author{H. D. Aller, M. F. Aller, \& P. A. Hughes}

   \institute{Department of Astronomy, University of Michigan, Ann Arbor,
      MI 48109 USA}

   \abstract{
      We interpret the linear polarization structures observed in
extragalactic radio sources, even those oriented at oblique angles to
the jet flows, to be due to oblique, relativistic shock fronts in the
emitting regions. Many sources exhibit indications of such oblique
structures, and the goal of this investigation is to test this
hypothesis quantitatively.  A selected group of ten highly-variable
extragalactic sources were observed with the VLBA at 15 and 43 GHz,
on nine epochs spanning a 30-month period; five of these objects were
also observed at 8.0 and 22 GHz. The integrated total flux densities
and linear polarizations of the selected objects were also observed
several times a month at 4.8, 8.0 and 14.5 GHz with the University of
Michigan 26-meter telescope. All objects exhibited variability with
several exhibiting more than one independent outburst during the period.
We show the evolution of the polarized components with time and discuss
the relativistic shock parameters required to match the observed
polarization structures. Even cases where the magnetic field is
apparently oriented along the jet flow can be fit by oblique shock
models when  relativistic aberration effects are included.     
}

   \maketitle
%

\section{Introduction}

There has been  considerable past success in fitting quantitative
transverse shock models to selected radio outbursts in extragalactic
sources where the radio polarized flux is parallel to the VLBI jet axis
(\cite{hughes89}).  However, many sources have exhibited bursts
where the radio polarization is at an arbitrary angle to the jet flow direction,
and  we suspect that in these cases, the linear polarization is produced by
{\it oblique} shocks.  To better understand the evolution of these oblique
polarization structures we  made multi-epoch observations of ten highly
active extragalactic sources selected from the  University of Michigan
variability program: DA~55, 0607-157, OJ~287, 1055+018, 3C~279, 1510-089,
OT~081, 1928+738, BL~Lacertae and 3C~446.  The goals of this study are to find
if, in fact, the observed polarizations can be produced by the alignment of
magnetic fields and acceleration of particles associated with oblique shocks in
relativistic jet flows, and to investigate what constraints can be placed
on the physical conditions and processes in the jets.  The interpretation is
made somewhat more complex by the effects of relativistic aberration.  Here
we report on preliminary results for one program source, BL Lacertae, based on 2~cm
measurements with the VLBA at five epochs and the monitoring results from the
University of Michigan ( hereafter UMRAO) 26-meter telescope.


\section{Results}
 
The single-antenna 14.5~GHz measurements of BL~Lacertae from the UMRAO are
shown in Figure~1.  Our observations at 4.8 and 8.0~GHz are not shown, to
simplify the figure, but they show similar variations.  A relatively large
outburst occurred in late 1999 through 2000, and there have been several smaller
flux events.  The epochs of the VLBA polarization observations are indicated by the
arrows along the abscissa in Figure 1, and the epochs of the five maps
shown in Figure 2. are designated
by the letters a to e.  The total polarized flux does not mimic the flux
density curve in a simple way.  The polarization position angle is consistently
offset from the average direction of the jet flow, $197^{\circ}$ (defined by
the direction from the core to the strongest jet component, and shown as the
horizontal line).  As discussed below, we believe that this offset is associated
with the curvature of the jet structure and the resulting orientation of
oblique shock structures to generate a deflection of the flow in the observed
direction.  

\begin{figure*}
\centering
   \vspace{400pt}
 \includegraphics{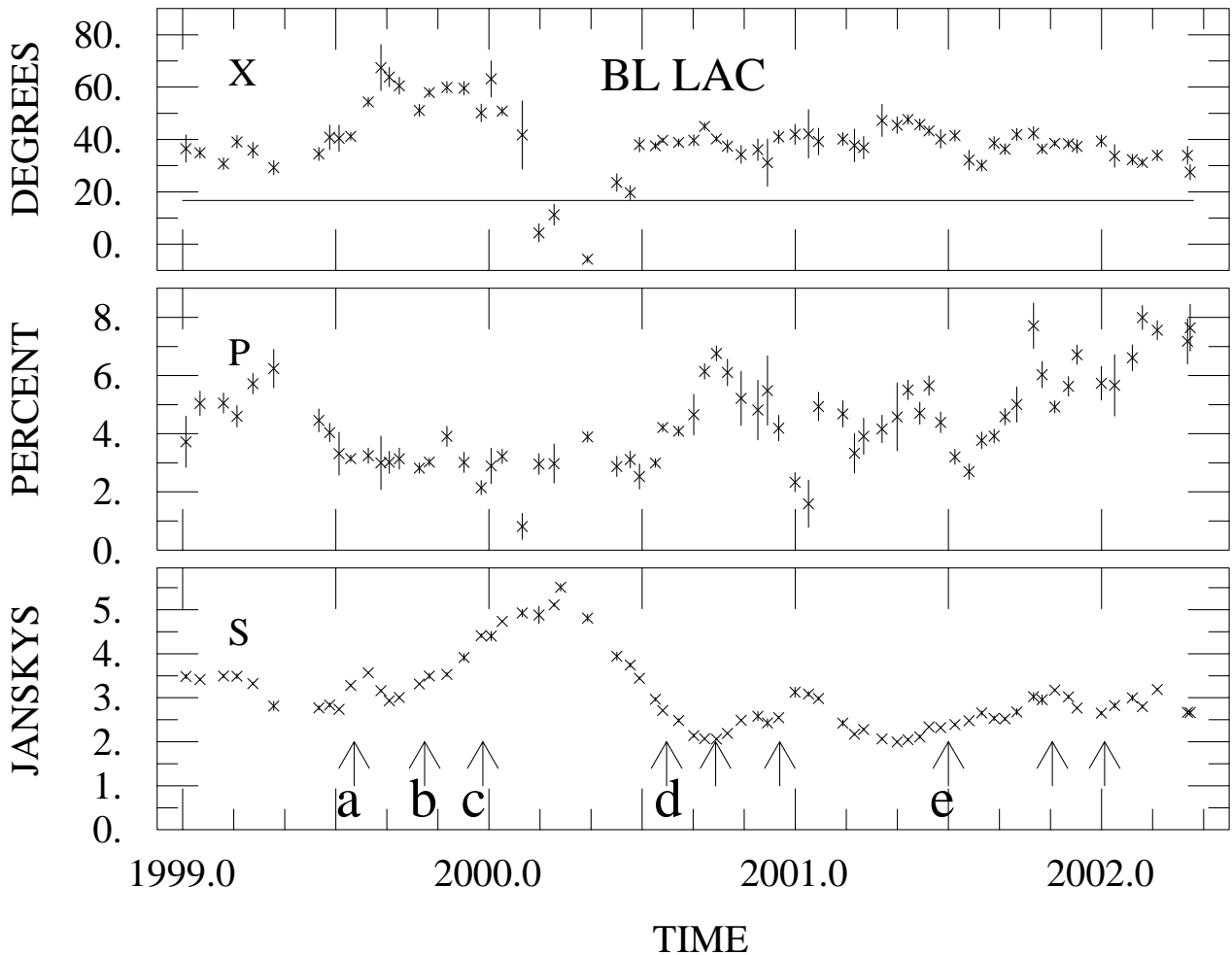}
 \caption{UMRAO light curves for BL~Lacertae at 14.5 GHz
based on two-week averages of the data. From bottom to top:
total flux density, fractional linear polarization, and position
angle of the electric vector of the polarized emission.
The polarization has been corrected for Faraday rotation assuming
a rotation measure of -205~rad/m$^2$.  The labeled arrows refer to 
the epochs of the images shown in Figure 2.
 \label{fig1}
  }
 \end{figure*}

 
  \begin{figure*}
   \centering
   \vspace{490pt}
   \includegraphics{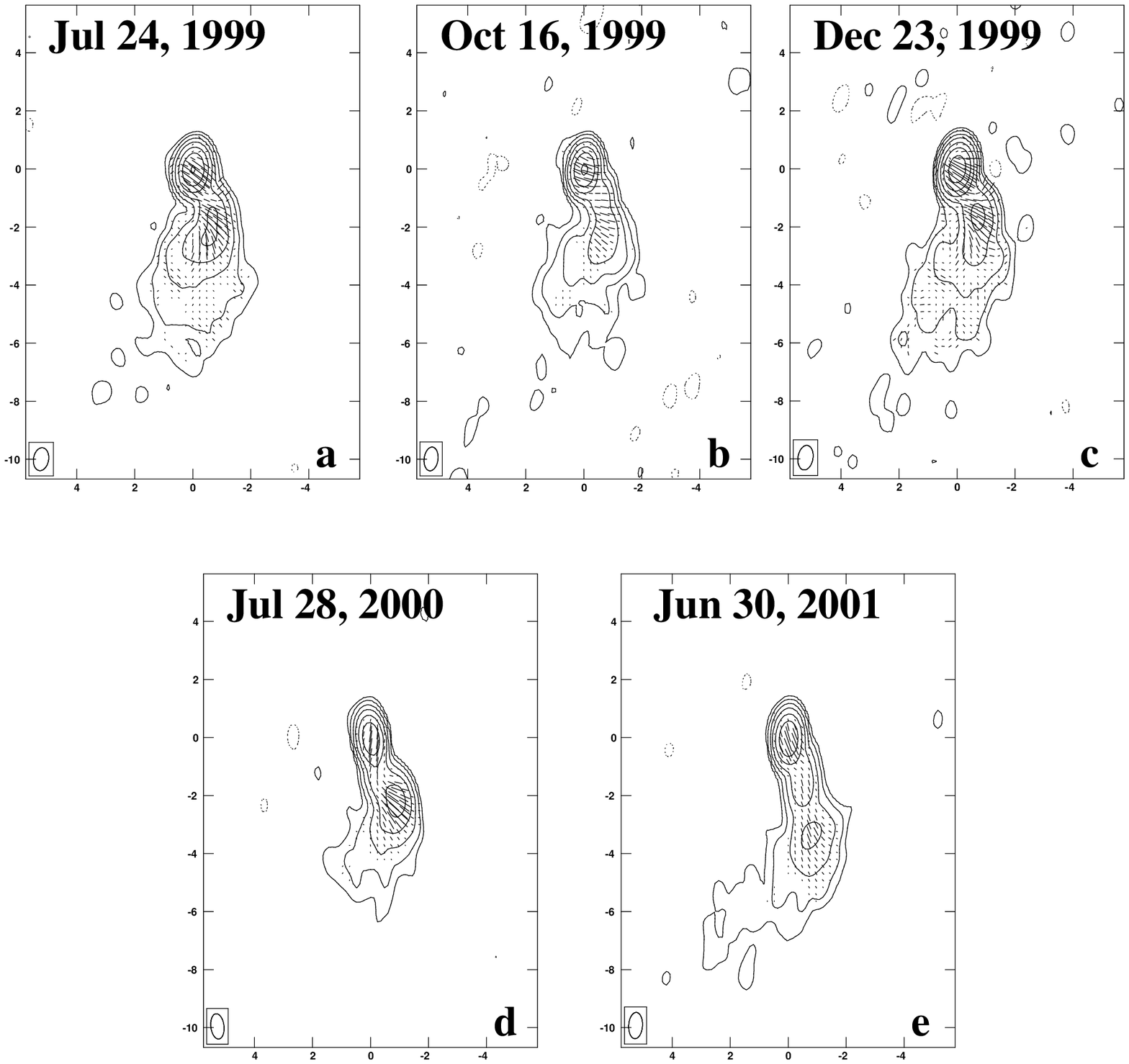}
\caption{VLBA images of BL Lacertae at 15.4~GHz at five epochs (shown in
Figure 1).  The contours show the intensity distributions on a logarithmic
scale; each level differs by a factor of square root of ten.  In each panel the
bottom contour represents 1.5 mJy/beam.  Dashed contours indicate regions
of negative flux in the maps.  The effective beam is shown in the lower
left of each image.  The lines superimposed on the contours show the
polarized flux and orientation of the electric vector: a line length of
1--mas equals 40 mJy/beam of polarized flux.
\label{fig2}
   }
 \end{figure*}

The five VLBA polarization maps are shown in Figure 2.  The
contour levels (chosen to emphasize the diffuse emission) are the same
in all images, but the noise levels vary due to weather conditions or
temporary equipment problems.  Note that the top three images (a, b,
and c) were obtained in a five-month period, while the last two images are
at a yearly spacing from the first image. The basic morphology of the
source remained unchanged over the two year observing period, and is
very similar to the structures observed by \cite{denn}.   The
`hot spot' evident in the jet appears to change position in an
erratic manner; it is not possible to define an `expansion rate'
from these images.  There is also considerable brightening and fading of the
outer  tail of the jet.  Although not evident without close inspection,
most of the  flux variations seen in Figure 1 are produced in the core
component; the first three epochs are associated with
the development of the burst that peaked in 2000, an event has almost
completely faded by epoch d. 

The majority of the variability (and the polarized flux) originates
in the unresolved (and part of the time partially opaque) `core'.
The most highly polarized component, however, is often the `hot spot'
located two or more mas from the core, where the degree of polarization is
as high as 25\%.  We believe that this is the site of a relatively
strong shock in the jet flow.  The striking feature is that this region
maintains a relatively fixed  polarization position angle (in the
vicinity of $45^{\circ}$) from epoch to epoch.  This is certainly not associated
with a transverse shock (such as fit the BL Lacertae data in the early
1980s).  The observed orientation is consistent with an oblique shock
which could be associated with the deflection or bending of the jet
flow towards the south east.
 
The description of an oblique shock in a relativistic flow is more
complex than for the case of a transverse shock.  We use the same
notation as \cite{cawth} who considered the case of conical shocks in
relativistic flows.  Their parameters, $\beta_{u}$ (the upstream
flow velocity), and the angles $\eta, \theta, \theta'$ and $\phi$ are
illustrated in Figure 3.  We have added an additional parameter (derivable
from the others) $\chi$, the flow deflection angle.  This angle can be
quite extensive as is illustrated in Figure 4.  Note that the deflection 
of the flow is in the opposite sense from the rotation of the EVPA
with respect to the initial flow direction.

\begin{figure}
\centering
\includegraphics[width=9cm]{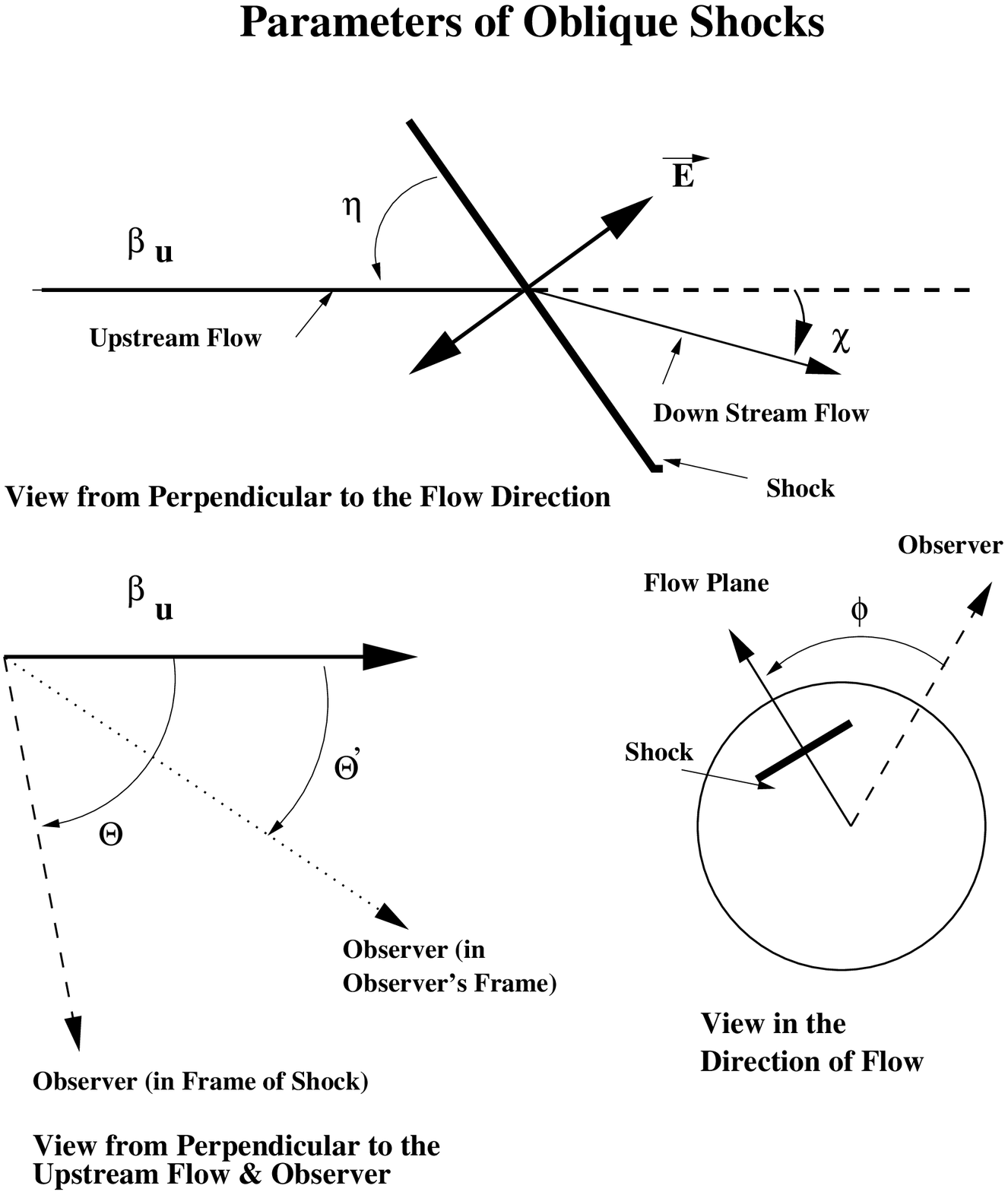}
\caption{Parameters needed to describe an oblique shock in a relativistic
flow (following \cite{cawth}). 
The strength of the shock (which sets the degree of compression
and hence the degree of polarization) is set by $\beta_{u}$ and $\eta$.
The observed polarization (orientation and degree of polarization) also
depend on the velocity of the shock, the angle to the observer, $\theta'$,
and the angle of the flow plane to the observer, $\phi$.
\label{fig3}
    }
\end{figure}

\begin{figure}
\centering
\includegraphics[angle=-90,width=9cm]{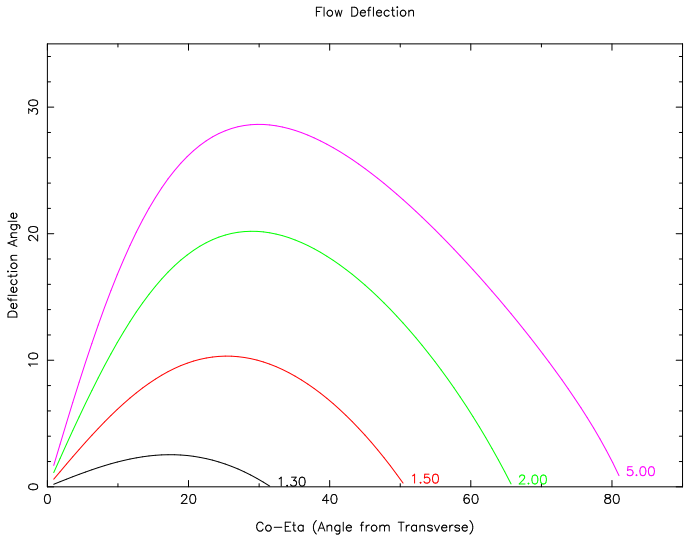}
\caption{The angle through which the flow is deflected at the shock
front, $\chi$, versus the strength of the shock (parameterized by the
Lorentz factor of the upstream flow) and the orientation of the shock
plane relative to the transverse direction.
\label{fig4}
   }
\end{figure}

We can define limits for the parameters of the shocks required
to reproduce individual maps, but there are many possible combinations of 
shock obliquity and observer viewiing angle that can produce the same apparent
polarization structure. Figure 5 illustrates the range of $\eta, \theta$, 
 and $\phi$ which will produce a 25\% polarized component at an
EVPA of $30^{\circ}$ to the initial flow direction.  
The deflection angles shown in Figure 4 are certainly large enough to 
generate the jet structures trailing off to the Southeast in Figure 2.
However, the eratic motion of the polarized peak, together with the 
unknown flow rate down the tail, makes it difficult to measure
the apparent deflection angle of the flow.  The erratic motion of the `hot spot', where
the deflection in flow direction takes place, may contribute to the widening
of the tail region, even if the deflection angle remains fixed.

\begin{figure}
\centering
\includegraphics[angle=-90,width=8.5cm]{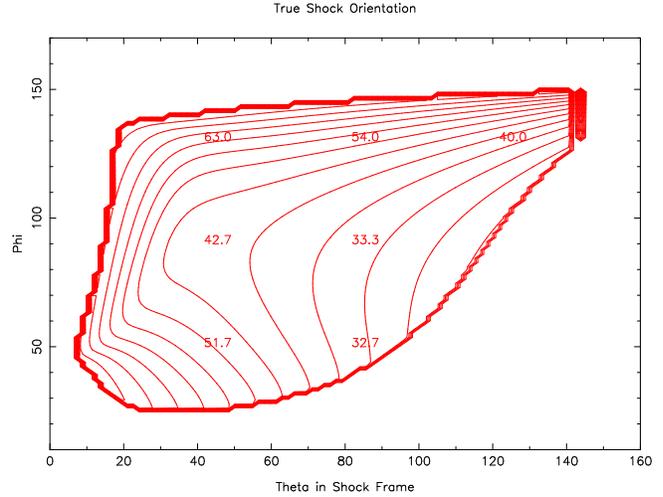}
\caption{Possible solution angles for the true angle of the shock front 
relative to the initial flow direction for an apparent polarization of
25\% at an EVPA (relative to the initial flow) of $30^{\circ}$. The contours
are at $5^{\circ}$ intervals.
\label{fig5}
   }
\end{figure}

\begin{figure}
\centering
\includegraphics[angle=-90,width=8.5cm]{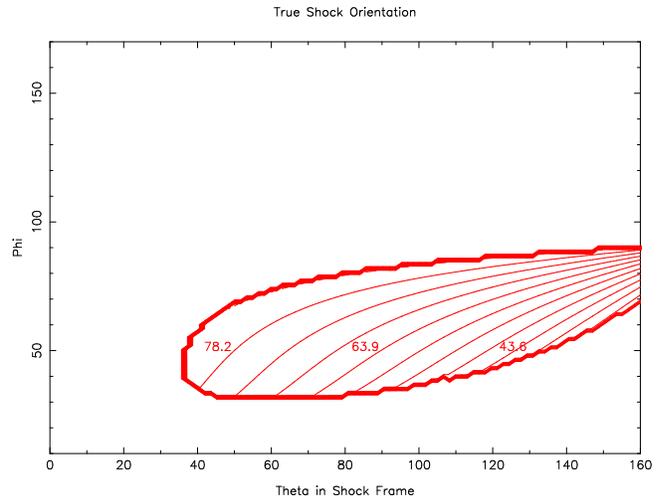}
\caption{Possible solution angles for the true angle of the shock front 
relative to the initial flow direction for an apparent polarization of
10\% at an EVPA (relative to the initial flow) of $90^{\circ}$. The contours
are at $5^{\circ}$ intervals.
\label{fig6}
   }
\end{figure}

Some regions of the BL Lacertae images presented here, and a dominant characteristic
of many QSO jets are polarization structures which are perpendicular to the flow direction.
Although this might be due to a shear layer, such structures can also be generated
by oblique shocks when the effects of relativistic aberration are taken into account.
Figure 6 shows the region of parameter space where an apparently perpendicular EVPA
could be generated. 


\section{Conclusions}

The polarized structures observed in the images that we have analysed
to date are all consistent with oblique shock structures which 
have similar shock strengths to those found
in previous studies of transverse shocks.  While the spacing of outbursts and
the time--scales of gross changes in the parsec scale jet structure 
of BL Lacertae are measured in years, images obtained months apart
reveal that there are significant changes in polarization structures
and erratic motions of intensity peaks on much shorter time scales.

\begin{acknowledgements}
This research was supported by grant AST-9900723 from the National
 Science Foundation. UMRAO is operated with funds from the
 Department of Astronomy of the University
of Michigan. The National Radio Astronomy Observatory is a facility
of the National Science Foundation operated under cooperative 
agreement by Associated Universities, Inc.

\end{acknowledgements}

\end{document}